\documentclass[12pt,preprint]{aastex}

\usepackage{natbib}

\newcommand{\be}{\begin{equation}}
\newcommand{\ee}{\end{equation}}
\newcommand{\nn}{\mbox{} \nonumber \\ \mbox{} }
\newcommand{\ba}{\begin{eqnarray}}
\newcommand{\ea}{\end{eqnarray}}
\newcommand{\om}{\omega}
\newcommand{\Alfven}{Alfv\'{e}n }

\newcommand{\E}{{\bf E}}
\newcommand{\B}{{\bf B}}
\newcommand{\J}{{\bf J}}
\renewcommand{\v}{{\bf v}}

\newcommand{\sech}{{\rm \,sech\,}}

\newcommand\eg{\textit{e.g.,\ }}

\newcommand{\Bf}{{magnetic field}}
\newcommand{\Bfs}{{magnetic fields}}

\newcommand{\NS}{neutron star}
\newcommand{\NSs}{{neutron stars}}
\newcommand{\EM}{electromagnetic}

\newcommand{\ms}{magnetosphere}

\begin{document}

\title{Magnetar activity mediated by  plastic deformations of \NS\ crust}

\author{Maxim Lyutikov\\
Department of Physics and Astronomy, Purdue University, 
 525 Northwestern Avenue,
West Lafayette, IN
47907-2036}

\begin{abstract}
We advance a "Solar flare" model of magnetar activity, whereas a slow  evolution of the magnetic field   in the upper crust,  driven by electron MHD (EMHD) flows,  twists the external magnetic  flux tubes, producing persistent emission, bursts and flares. At the same time the \NS\ crust  plastically  relieves the imposed \Bf\ stress,    limiting  the strain  $ \epsilon_t $ to values well below the critical strain $ \epsilon_{crit}$ of a brittle fracture,  $ \epsilon_t \sim 10^{-2}\epsilon_{crit} $.

Magnetar-like behavior,   occurring  near the  magnetic equator, takes place in all \NSs, but to a different extent.  The persistent  luminosity is proportional to cubic power of the  \Bf\  (at a given age), and hence is hardly  observable in most rotationally powered \NSs. 
Giant flares can occur only  if the \Bf\ exceeds some threshold value, while 
 smaller bursts and  flares may take place in relatively small \Bfs.

Bursts and flares are   magnetospheric reconnection events  that launch \Alfven shocks which convert into  high frequency whistlers upon hitting  the \NS\ surface. The resulting whistler pulse induces a strain that increases with depth both due to the increasing electron density (and the resulting slowing of the waves), and due to the increasing coherence of a  whistler pulse with depth. The whistler pulse is   dissipated  on a time scale of approximately a  day at shallow depths corresponding to $\rho \sim 10^{10} {\rm g cm}^{-3}$; this  energy is detected as  enhanced post-flare surface emission.

 \end{abstract}

\maketitle

 \section{Two competing models of magnetar bursts and  flares}

 Two closely related classes of young neutron stars -- Anomalous X-ray Pulsars
(AXPs) and the Soft Gamma-ray Repeaters (SGRs) -- both show X-ray flares and   quiescent X-ray emission \citep[see][for review]{2006csxs.book..547W}.  Their high energy emission is powered by dissipation of super-strong \Bfs, $B > 10^{15}$G \citep{TD95}. 

Two models of magnetar  bursts and flares have been proposed. In the first, a star-quake model, a
 flare relies on a {\it sudden}  fracture of the \NS\ crust that lead to fast, on time scale of hundred milliseconds, untwisting of the internal
magnetic field and a resulting twisting-up of the external magnetic
field \citep{TD95}.  Alternatively,   a Solar paradigm postulates that  {\it slow} evolution  of the internal magnetic
field leads to a gradual twisting of magnetospheric field lines on a
time scale much longer than that of the giant flare \citep{LyutikovTear,lyut06}.  Eventually, the increasing
twist  associated with the current-carrying magnetic field in the magnetosphere lead to  a  sudden relaxation of the external \Bf, accompanied by associated
dissipation and change of magnetic topology. 

Perhaps the best observational  argument in favor of external magnetic dissipation during the burst is  the observed sharp rise
of $\gamma$-ray flux during giant flares,  on the time scale similar to the \Alfven crossing time of the inner
magnetosphere, $\sim 0.25 $ msec \citep[][]{2005Natur.434.1107P}. This
 points to the magnetospheric
origin of giant flares \citep{lyut06}.

The efficient crack formation, needed in the star-quake model, requires the solid's ability to form a small void at the crack's
location.  Since in the neutron-star crust the pressure is greater than the shear
modulus by two orders of magnitude, the conventional crack that relies on a formation of the void cannot occur \citep{2003ApJ...595..342J} \citep[though so called  deep Earth quakes][apparently violate this condition]{Frohlich}. 
Secondly, even if elastic properties of the crust did allow brittle cracking, the energy release would  be strongly suppressed by the  magnetic tension (even though the \Bf\ itself  leads to cracking!).
\cite{2012MNRAS.427.1574L} demonstrated that the energy release from a thin crack would be strongly suppressed, since the magnetic field provides mechanical connection between the two slipping sides of the crack and rapidly suppresses the slippage. The key point is that magnetically induced cracks form orthogonally to the \Bf; then the   tension
strongly suppresses  the slippage and the energy release.  Thin cracks release energy 
 via magnetic-field diffusion too slowly to be able to contribute to the
energetics of magnetar flares.  In addition, the waves generated within \NS\ experience many  internal reflections due to impedance mismatch. As a result,  the rise time of the emission is  minutes to hours, too long account for $\sim$ millisecond rise time of bursts and  flares \citep{2013arXiv1312.5144L}.
Thus, {\it  even if mechanical properties of the crust allowed brittle fracture, the resulting energy release on the time scale of flares is very small.}

 \section{Evolution of \Bf\ in \NSs}
\label{evolution}

 \subsection{Generation and  initial relaxation of \Bf}
 
 We envision the following paradigm for evolution of \Bf\ in \NS\ crusts, and specifically  in magnetars.
Strong magnetic fields, of the order of $10^{16}$ G in case of magnetars,
may be created  by a
dynamo mechanism, \eg\ of the $\alpha-\Omega$ type,  operating at birth of neutron stars  \citep{TD93}. It is expected that the dynamo operates most efficiently in the outer layers of the proto-neutron star, where  neutrino-driven turbulence is most efficient  \citep{1995PhR...256....5C,1994ApJ...435..339H}.

Typically, neutrino driven turbulence dies out on time scales of seconds,  well before the crust solidifies (though some fluid motion, like a shearing rotation, may persist for longer times).
As a result,  during the time that the star remains fluid the \Bf\ reaches an  MHD-type equilibrium.
In   MHD equilibria  Lorentz  forces are balanced by gradients of pressure and gravitational forces,
\be
\J \times \B =  \nabla p + \rho \nabla \Phi,
\label{MHD}
\ee where $\Phi$ is the gravitational potential.
By dividing Eq. (\ref{MHD}) by $\rho$ and  taking a  curl, we find \citep[\eg][]{Reisenegger07}
 \be 
\nabla \times  {\J \times \B  \over \rho}= - {\nabla p \times \nabla \rho \over \rho^2} .
\label{MHD2}
\ee

Stability  of fluid stars requires that the configuration  involves both toroidal and poloidal \Bfs\ \citep{FlowersRuderman,BraithwaiteSpruit}. In addition, it appear that purely barotropic stars are unstable \citep{2012MNRAS.424..482L}. 
In stable configurations the energy of the toroidal \Bf\ is {\it at least} comparable to the energy of the poloidal field, and may greatly exceed it \citep{2013MNRAS.433.2445A}. In addition, toroidal \Bf\ are confined to relatively small volume near the surface at the magnetic equator. As a result, the value of the toroidal \Bf\ may exceed the poloidal by as much as two orders of magnitude, reaching  as high as $B \sim 10^{16} $ G \citep{2013MNRAS.433.2445A}.

\subsection{Freezing and reviving the magnetic  dynamics: electron MHD}

At approximately 100 seconds the crust cools sufficiently and ions in   the
neutron star crusts  form a fixed lattice, while for slow perturbations  the electrons behave as an  inertialess fluid. The resulting dynamical system is called  electron  MHD  (EMHD) \citep{1960RSPTA.252..397L,Kingsep,1994PhR...243..215G,RG}. Within the framework of EMHD, the \Bf\ is frozen into electron fluid. In the limit of infinite conductivity,   the electric field  then satisfies the condition
\be
\E+\v_e \times\B=0,
\ee
and  the induction equation becomes
\be
{\partial\B\over\partial t} =  \nabla \times    \left(\v_e  \times \B \right).
\label{induction}
\ee

For infinitely rigid lattice the 
 the electric current is  produced by the flow of  electrons, while ions provide a neutralizing background. In case of plastic deformations the flowing ions also provide current density, so that   $\J=ne(\v_i - \v_e)$, where $n$, $-e$, and $\v_e $ and $\v_i $ are the local number density, charge, and
average velocity of the electrons. As we demonstrate below, see Eq. (\ref{main13}), for the chosen model of plastic deformations  the ion velocity is typically much smaller than the electron velocity. Neglecting the contribution of ions to the current density,
$\J=-ne\v_e ={c\over 4\pi}\nabla\times\B$, the only dynamical variable in this equation is the magnetic field   \citep{Kingsep}:
\be
{\partial\B\over\partial t} = - {c \over 4 \pi  e} \nabla \times    \left({\nabla  \times \B  \over n} \times \B \right)
\label{main}
\ee
(density $n$ is an externally prescribed function).
 So, given an initial field configuration and appropriate
boundary conditions, the induction equation uniquely determines its evolution.

After the crust freezes, it  no longer obeys MHD, but EMHD equations. The initial field, satisfying the MHD equilibrium condition, Eq.~(\ref{MHD2}), starts evolving according to Eq.~(\ref{main}), which initially (replacing Eq.~[\ref{MHD}] in it) takes the form
\be
{\partial\B\over\partial t}=-{1\over e}\left(\nabla{1\over n}\times\nabla p+\nabla{\rho\over n}\times\nabla\Phi\right).
\label{EarlyEvol}
\ee
In the initial MHD equilibrium, due to the Lorentz forces and the stable, compositional stratification of the neutron-star matter \citep{RG}, the gradients in this equation are not exactly radial and not exactly parallel to each other, although their relative angles are generally small, $\sim B^2/(8\pi p)\leq 1$ \citep{Reisenegger07}.
Thus, {\it  an MHD-equilibrium \Bf\ structure is, generally, not an equilibrium configuration of EMHD. }
Two terms will drive evolution: the non-barotropic
condition $\nabla p \times \nabla \rho \neq 0$ and the stratification of the \NS\ material, $\rho / n \neq$constant.

There are exception to the above (that freezing of MHD equilibrium results in non-equilibrium EMHD state): (i) if MHD configuration is force-free, (ii) if the fluid is
 barotropic, when $p=p(\rho)$ and has constant mean molecular weight (or constant proton fraction in case of {\NS}s), $\rho/n=$ const, (iii) if the system is effectively one-dimensional, will all the quantities depending only on one particular coordinate. In these cases
 MHD equilibrium corresponds to EMHD equilibrium as well.
We do not expect that any of the three condition listed above are satisfied in proto-neutron stars. Thus,
after the crust freezes the resulting \Bf\ state is, generally, not in Hall equilibrium, and  a system start dynamical evolution. Unfortunately, we cannot give an explicit example of this very fundamental statement, since no analytical examples of at least two-dimensional stable MHD equilibria in stars exist.

  The normal modes of EMHD plasma are  whistlers. Harmonic whistlers of {\it  arbitrary}  amplitude $ \delta B$  are   exact solution of EMHD  \citep{2013PhRvE..88e3103L},
\ba && 
\B=B_0  {\bf e} _z+ \delta B {\bf e}_B e^{-i (\om t - k (z \cos \theta + x \sin \theta))}
\nn &&
\om  = c^2 k^2 | \cos \theta |  {\om_B / \om_p^2}
\nn && 
{\bf e}_B={1\over 2} \,  \{  i \cos \theta, 1 , - i  \sin \theta \} ,
\label{whistlers}
\ea
where  $\om_B$ and $\om_p$ are  electron cyclotron and plasma frequencies,  and ${\bf e}_B$ is  the eigenvector.

A whistler wave with  fluctuating \Bf\  $\delta B$ induces stresses in the crust of the order of $\sigma \sim (\delta B /B_0) B_0^2$. This stress is initially zero and builds up to this value on the Hall time scale \citep{RG}
\be
\tau_H = { L^2 \om_p^2 \over c^2  \om_B} \approx  100\,  {\rm yrs} \left({L \over 1\,  {\rm km}}\right)^2 \rho_{10} b_{15}^{-1}
\label{tH}
 \, {\rm yr},
\ee
where $L \sim 1 {\rm km} $ is a typical scale of \Bf\ fluctuation (crust thickness) and $\rho_{10} = \rho/(10^{10}) {\rm g cm^{-3}}$ is typical  density at the neutron drip point, $b_{15} = B/(10^{15} {\rm G})$ . The Hall time scale has a steep dependence on the scale, density and \Bf, and can vary considerably within the crust.

Thus,   freezing of MHD equilibrium results in an non-equilibrium EMHD state, which revives the evolution of the \Bf\ and 
 leads to generation of whistler waves. 
At the moment of crust solidification, there are no shear stresses, but as the \Bf\ evolves on the Hall time scale,  shear stresses build-up,  reaching a maximum value on the time scale of order of the crust Hall time. 

The initial MHD-stable state is dominated by low multipoles (\eg\ axisymmetric dipole-like configuration of \cite{BraithwaiteSpruit}). The evolution of the \Bf\ in the crusts then proceeds to  a number of processes: (i) turbulent EMHD cascade, that creates higher multipole \citep{1973JETP...37...73V,1999PhPl....6..751B,2003PhPl...10.3065G,2004ApJ...615L..41C,  2008PhRvL.100f5004H,2013arXiv1306.4544L}; (ii)  large scale Hall drifts  that may lead to the formation of current sheets approximately in one Hall time \citep{Kingsep,1994PhR...243..215G}; (iii) crustal yields that dissipate the energy of whistler waves. All this effects proceed, roughly on a Hall time scale. It is not clear at the moment which of the above  is the dominant mechanism of the \Bf\ evolution. 

\subsection{Crustal plastic properties do not determine magnetar activity}

Importantly, the evolution of the \Bf\ in the crust proceeds in the electron MHD (EMHD) regime, where the \Bf\ is frozen in the electron fluid. This may lead to the evolution of the external fields and externally-generated flares {\it even without any crustal yielding}, be it ductile or brittle. The crust may still respond to the imposed stress (\eg\ relieving the stress plastically, as we discuss in this paper), but it does not need to in order to produce a flare: even if the crust were absolutely rigid the crustal  \Bf\ would  evolve (and twist the external \Bf)   due to the EMHD drift. 

On the other hand, the post-burst enhanced surface thermal emission does indicate substantial energy deposition deep within the crust during the bursts and flares \citep{2006astro.ph.11747E,2012ApJ...761...66S}. \cite{2012ApJ...761...66S} inferred the deposition density of $\sim 10^{10} $  g cm$^{-3}$, close to the neutron drip line, the outer core boundary located approximately hundred meters below the surface. One of the points of the paper is to reconcile the external energy deposition and  observation of a long-term (weeks to months) post-burst surface  cooling.

 \section{A model problem: plastic deformations of the crust}

Evolution of the \Bf\ due to Hall drift of the electron fluid  induces shear stresses in the ion lattice, which may yield.   Thus, the crustal dynamics is described by two separate processes: Hall electron drifts and the response of the crust (which can take various forms:  plastic, ductile or brittle).

In the NS crust the shear modulus increases monotonically with depth, and then quickly  goes to zero once the crust dissolves. The shear modulus changes according to \citep{1973NuPhA.207..298N,TD01}
\be
\mu \approx 6 \times 10^{24}\,  \rho_{10}^{0.8}\, {\rm erg \,  cm}^{-3} .
\ee
We expect that when the \Bf\ energy density exceeds  the shear modulus, 
 for fields 
\be
B >B_\mu= \sqrt{4 \pi \mu} \sim 9 \times 10^{13} \,   \rho_{10}^{0.4}  \, {\rm G},
\ee
the crust {\it must} respond plastically (but,  it  can also respond plastically at much smaller \Bfs). In fact, plastic response may dominate the lattice response even for stresses much smaller than the shear modus, especially for   slow strain rates.


 In this paper we consider  plastic  response of the crust given an externally  imposed  Hall magnetic stresses,  from the \Bf\ frozen in the evolving electron fluid.  
 The typical time scale of the  electron drift, the Hall time (\ref{tH}), is much  longer compared to the observed duration of the magnetars'  bursts and flares, so we can assume that the stress builds slowly, quasi-statically. The normal modes of the Hall plasma - whistlers - carry a shear stress, assumed to be balanced by the lattice stress. As a simplified model of this whistler-carried stress on the ion lattice, we assumed that  whistlers create a simple two-dimensional current layer where the  $z$-displacement of the electron fluid ${\xi_e}$ evolves according to 
 \be
 \dot{\xi_e} =(\xi_0 /t_H) \tanh( x/a), 
 \ee
 where  $\xi_0$ is the typical amplitude of the lateral displacement of the electron fluid, $a$ is a typical thickness of the current layer  (realistically, $\xi_0 \sim a$) and $t_H$ is a typical time of the growth of the electron stress on the ion lattice.
The strain in the electron fluid ${\epsilon}_e$  then is growing linearly in time according to the law
 \be
 \dot{\epsilon}_e = {\xi_0 \over a} {1\over t_H} \cosh^{-2} (x/a),
 \label{ee}
 \ee 
 where dot denotes time derivative.  The parallel \Bf, frozen in the electron fluid, then evolves according to  
\ba &&
B_z = B_0 \partial _x \xi_e(t) =  B_0 \epsilon_e
\nn &&
 \epsilon_e ={ \xi_0\over a}  {t \over t_H}  \sech^{2} (x/a)
 \ea
 
 The \Bf\  exerts a stress on the ion lattice, that results in an elastic strain rate
 \be
\dot{  \epsilon} _{el} = {B_0 \dot{B_z}  \over 4 \pi \mu} = {v_A^2 \over c_s^2} { \dot{B_z}  \over B_0}  = \dot{ \epsilon_e} {v_A^2 \over c_s^2},
  \ee
  where $v_A$ is \Alfven velocity and $c_s$ is shear speed.
   We assume that the ion lattice  responds in a viscoelastic  way  to the stress induced by the Hall motions of electrons (\ref{ee}).   Thus, in addition to the elastic strain rate there is also a plastic relaxation rate. 
   
  Plastic deformations in crystals are typically divided into two categories. First, there is plasticity that occurs beyond some critical strain; below this critical stain there are no irreversible deformations. Secondly, plastic deformations can occur at small strains. This type of plastic deformations is called creep (or, related, a plastic relaxation) - in this case even small stress leads to irreversible deformations (like in a pure unseeded aluminum). Generally, all  materials exhibit some viscoelastic response, while the existence of a well-defined yield stress is the exception rather than the rule \citep{Lublinerbook}.

 Plastic creep/relaxation in crystals is induced by the motions of crystal deformations \citep{1934ZPhy...89..605O,Gilman}. Shear deformations are due to the motions of the line defects.
 The plastic rate of strain depends on the surface density of defects $\rho_d$ (dimensions cm$^{-2}$), the rate of the defects's motion $v_d$ and a typical distance a defect travels in one jump (the Burgers vector) $d$ via the Orowan equation
 \be
 \dot{\epsilon}_p = - d \rho_d v_d
 \label{ep}
 \ee
 (the minus sign reflects  the fact that the rate of the plastic stress, like viscosity, works as  a damper.)
 The length of the Burgers vector is of the order of the inter-particle distance, $d \sim n^{-1/3}$, $n = \rho /(A m_p)$, where $A$ is the mass atomic number of the ion lattice. 
  
 The rates of strain of the ion lattice due to the EMHD dynamics and plastic relaxation add:
  \be
  \dot{\epsilon}= 
 \dot{\epsilon}_p+ \dot{\epsilon}_{el} .
 \label{mm}
 \ee
 Eq. (\ref{mm})  describes  viscoelastic  deformation due to externally imposed, time-dependent strain ${\epsilon}_{el} $ and the plastic relaxation ${\epsilon}_{p} $.  

The density of the dislocations $\rho_d$ and the defect drift velocity $v_d$ are, generally, functions of the stress, activation energy of the defect's movement, temperature and history of  stress \cite[\eg][]{Gilman} -- 
determining the density of dislocation $\rho$ and their drift velocity $v_d$ is a complicated problem. Typically, at higher temperatures the density and especially  the mobility of dislocations obey an Arrhenius-type scaling, $\propto \exp\{ - \Delta E /T \}$, where $ \Delta E$ is an activation energy. On the other hand, at small temperatures/high stresses the mobility and the density of dislocations are determined by the applied stress. 
In this case the density of dislocations  may be estimated as \citep{Gilman}
 \be
 \rho_d = \left( { \epsilon\over d} \right)^2
 \ee
 (this scaling uses the fact that a stress from a dislocation line varies with distance  according to $\sigma \approx \mu  (d/r)$, so a mean stress is $\sim  \mu d \sqrt{\rho_d}$).

We parametrize the dislocations' velocity   as \citep{1960AuJPh..13..327G}
 \be
 v=c_s e^{-\epsilon_{crit}/\epsilon},
 \ee
where $\epsilon_{crit}$ is the critical fracture strain. (So that at the critical strain the dislocation velocity is a fraction  $1/e$ of the shear speed). 

 The total rate of strain of the lattice  is given by the EMHD strain (\ref{ee}) and the plastic strain (\ref{ep})
 \be
    \dot{\epsilon} = - {c_s \over d} {1\over \epsilon^2} e^{-\epsilon_{crit}/\epsilon} + {\xi_0 \over a} {1\over t_H}  {v_A^2 \over c_s^2} \cosh^{-2} (x/a).
    \label{main11}
    \ee
 Equation (\ref{main11}) represents a balance of the EMHD stress and the plastic stress on the lattice, it  determines the  evolution of the strain $\epsilon$ as a function of time and coordinate $x$ (as a parameter) and properties of the medium. The lattice  strain  induced by the motion of the electron fluid is relieved via  creep.  
 
 Beyond the Maxwell time, Eq. (\ref{tm}), the flow reaches a steady state, $\dot{\epsilon}=0$   \footnote{The steady state is reached only for sufficiently slow driving since the plastic strain rate, $ \propto e^{-\epsilon_{crit}/\epsilon} /  \epsilon^2$, has a maximum value of $\approx 0.5 / \epsilon_{crit}^2$ reached at $\epsilon = \epsilon_{crit}/2$. So, for $t_H \leq 3.4\epsilon_{crit}^4  (b/c_s) (\xi_0/a)  (v_A^2 / c_s^2)$ plastic relaxation cannot compensate for the externally imposed shear.} with the terminal strain (up to logarithmic accuracy), 
 \be
 \epsilon_t \approx   {\epsilon_{crit} \over \ln \left({ a \over  \xi_0}{ c_s t_H   \over  d } { c_s^2\over  v_A^2} \sech^2 (x/a)\right)} \approx    {\epsilon_{crit} \over \ln \left({  c_s t_H   \over  d }\right) }.
 \label{t}
 \ee
 The argument of the logarithm in (\ref{t}) is very large,  the shear velocity times  the Hall time divided by the  inter-particle distance $d$. There is also a weak, logarithmic dependence on the parameter $\xi_0/a$, the ratio of the lateral deformation of the electron fluid to the thickness of the current layer and on the ratio of shear speed to \Alfven speed $c_s/v_A$.
 Typically, for  parameters in  the NS crust, 
\be
 \epsilon_t \approx (1-  3) \times 10^{-2} \epsilon_{crit}  \approx 10^{-4},
 \label{t1}
 \ee 
 where we adopted $\epsilon_{crit} \approx 10^{-2}$ for the critical strain \citep{2012MNRAS.426.2404H}, \citep[see also][who argue for larger   $\epsilon_{crit} \approx 10^{-1}$]{2009PhRvL.102s1102H}.
 
  A time to reach the steady state (Maxwell time) is
  \be
  t_M \approx { a \over \xi_0} \epsilon_t  t_H \approx \epsilon_t  t_H  \approx  {\rm   months - years}
  \label{tm}
  \ee
  (we assumed that the lateral $ \xi_0$ displacement of the  plastically flowing layer is of the order of  its thickness $a$.)

  Thus, after a fairly short time a plastic creep relieves the crustal strain at a level of a few percent of the critical strain. At the same time the assumed deformations of  the electron fluid proceed linearly in time, increasing on the Hall time scale. Under the EMHD approximation the \Bf\ is frozen in the electron fluid, so that the deformations of the \Bf\ increase linearly with time (recall, we assume that the plasma is ideal, so there is no slipping of the \Bf\ through the electron fluid). The deformations of the \Bf\ induced by the motion of the electron fluid within the crust will lead to the similar deformations of the \Bf\ outside the \NS\ -- twisting of the external \Bf.


\begin{figure}[h!]
\includegraphics[width=.49\linewidth]{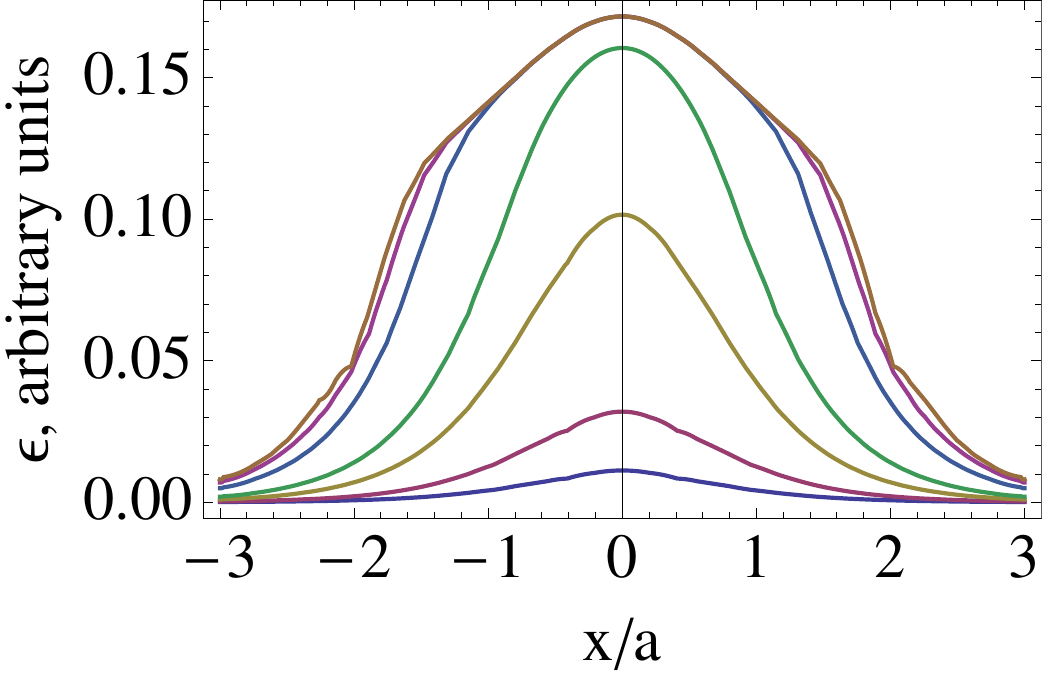}
\includegraphics[width=.49\linewidth]{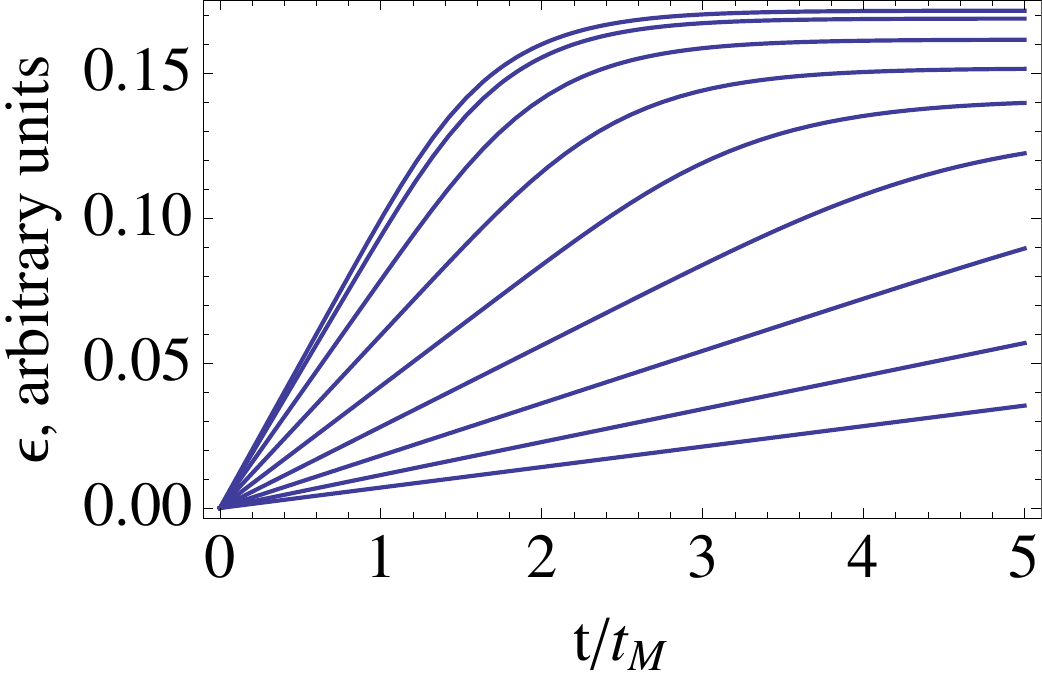}
\caption{ Evolution of the strain in the plastically relaxing medium subject to the externally-imposed stress $\propto t  \sech^2 (x/a)$ (linearly increasing with time). {\it Left Panel:} strain as function of $x$ at different times; {\it Right Panel:} strain as function of time  at  different $x$ (top to bottom curves, $x$ increases by $0.25 a $).   Initial strain is zero; at early times the response of the medium is mostly elastic, balancing the externally-imposed stress.  As time progresses,  the system reaches a steady state, balancing the  rates  of the elastic strain and  plastic relaxation. At smaller $x$, where the strain grows quicker, the steady state is reached earlier in time. For the chosen driving, the terminal shape is, approximately, $\epsilon \propto ({\rm Const}+ \ln \left\{  \sech^2 (x/a)\right\} )^{-1}$, Eq. (\ref{t}). 
}
\label{eofXT}
\end{figure}


Lorentz stresses lead to the distortion of the crustal lattice and, thus, do work on the crystal (this is a deviation from the EMHD, where the lattice is assumed to be infinitely ridged). At the steady state all the work done by the \Bf\ is dissipated plastically, so that the 
  plastically dissipated energy  is 
\be
W=  \sigma_{el}  {v_A^2 \over c_s^2} \dot{ \epsilon_e} 
\approx
 \epsilon_{t}  \dot{U}_B
 \approx 10^{-4} {U}_B/ t_H,
 \label{W}
 \ee
 where $U_B = B^2/(8 \pi)$ is the magnetic energy density.

Next, let us verify that the velocity of ions due to plastic deformations is much smaller than the Hall drift velocity of electrons (which is $\sim a/t_H$).  The typical velocity of the ion component can be estimated as, see Eq. (\ref{main11})
 \be
 v_i \sim a    \dot{\epsilon} =  {c_s a \over d} {1\over \epsilon^2} e^{-\epsilon_{crit}/\epsilon} \approx {a \over t_H} \, {1\over \epsilon_t^2} 
 \left({c_s t_H \over b} \right)^{-1/\epsilon_{crit}},
    \label{main13}
    \ee
    where we used Eq. (\ref{t}) to estimate the typical strain.  Since the term  $ c_s t_H / b$  is logarithmically large, and is raised to large negative power, the ion velocity 
  can be neglected.

Finally, let us comment on the possibility of crust melting during propagation of crustal  faults  \citep{2014arXiv1406.4850B}.  The plastic flowing layer layer  is heated viscously, while  cools by conduction. Can the resulting dissipation lead to local melting of the layer?  The answer critically depends on whether an initially distributed stress localizes into the narrow bands. At higher temperatures the effective viscosity is smaller; this could affect the overall evolution of the crust via so called shear thinning. If there is no localization of stress,  the answer to the above question is clearly not.   Balancing energy generation rate, $\sim W a$,  and the heat  flow per unit area of the crack, of the order of $\kappa_T T_{\rm melt} /a$, where $\kappa_T=  10^{20} {\rm erg\,  cm^{-1} s^{-1} K^{-1}}$ is thermal conductivity and $T_{\rm melt} \approx 10^9 {\rm K},$ is melting temperature  \citep{1976ApJ...206..218F,1979ApJ...230..847F,1981ApJ...250..750F,1999A&A...346..345P,2012arXiv1201.5602P}, the required size of the plastically flowing layer,
\be
a_{\rm melt} = \left( {\kappa_T T_{\rm melt} t_H   \over B_0^2 \epsilon_{t}  } \right)^{1/2} \sim  10^8 t_{H,4} b_{14}^{-1}\, {\rm cm},
\ee
is much larger than the size of the \NS.

On the other hand, some material do show stress localization - formation of so-called shear bands  \citep{2002pmas.book.....W}. This  dynamic localization of stress (as opposed to stress localization due to preexisting defects) depends critically on the details of the shape of the stress-strain curve \citep{1975JMPSo..23..371R,2002JGRB..107.2045M}, which are not known for the \NS\ crusts.  Typically, stress localization occurs for {\it  compressive} deformations \citep[as opposed to shear deformations][]{Rice76thelocalization};  compressive deformations are not likely to be important in \NS\ crusts.

 \section{Evolution of the external twist}

 \subsection{Persistent emission}
 
 Motion of the electron fluid inside the crust twists the outside \Bf\ - the electric current is pushed outside of the \NS. As the current-carrying particles hit the surface, they are stopped by the interaction with the crust, heating the surface -  this plays a role of resistivity, leading to continuos untwisting of the magnetic flux. In order to produce flares through instabilities of the twisted flux tube the rate of twist, of the order of the Hall time scale,  should exceed the rate of this resistive relaxation. 
 
 Consider a magnetic flux tube that at the stellar surface has a radius $r_{t,0}$ and carries total current $I = e n c \pi r_{t,0}^2$ (assuming that the charge carries move relativistically). By flux conservation,  the radius of the flux tube $r_t$ changes with height according to $r_t =  r_{t,0} (r/R_{NS})^{3/2}$. Typical toroidal \Bf\ within the flux tube then changes according to
 \be
 B_\phi = {I \over 2 \pi c r_{t,0}} (r/R_{NS})^{-3/2}.
 \ee
 The field line twist changes with height according to
 \be
 \Delta \phi = { B_\phi  \over B_p} =(r/R_{NS})^{3/2} {I \over 2 \pi  c B_{NS}   r_{t,0}} =  \Delta \phi _{\rm max}( r /r_{\rm max})^{3/2} 
 \ee 
 The maximal twist of the \Bf\  line is reached at the highest point $r_{\rm max}$:
 \be
 \Delta \phi _{\rm max}=  {I \over 2 \pi c B_{NS} r_{t,0}} (r_{\rm max}/R_{NS})^{3/2}.
 \label{phi}
 \ee
 For  a stable flux tube it is required that  $ \Delta \phi _{\rm max} \leq 1$.
 
 We can use Eq. (\ref{phi}) to parametrize the current in term of the maximal height reached by the flux tube and the maximal twist
 \be
 I = { 2 \pi c} B_{NS} r_{t,0} (r_{\rm max}/R_{NS})^{-3/2}  \Delta \phi _{\rm max}.
 \label{I}
 \ee
 Thus,
 \be
 B_\phi= { R_{NS}^3 \over r^{3/2} r_{\rm max}^{3/2}} \Delta \phi _{\rm max}  B_{NS}.
 \ee
 
 Note, that the energy stored in the toroidal \Bf\ per unit $dr$ is independent of radius
 \be
 {B_\phi^2 \over 8 \pi}  \pi r_t^2 dr = \Delta \phi _{\rm max}^2  (r_{\rm max}/R_{NS})^{-3}  { B_{NS} ^2 r_{t,0} ^2 \over 8} dr.
 \ee
 (For a given maximal twist $ \Delta \phi _{\rm max}$ and given radius at the \NS\ surface  the  flux tubes that extend to larger heights carry less toroidal energy.)
 The total  energy in the toroidal \Bf\  can be estimated as 
 \be
 U_B =  \Delta \phi _{\rm max}^2   { B_{NS} ^2 r_{t,0} ^2  R_{NS}^{3}  \over 8 r_{\rm max}^2}.
 \ee
 
 If current-carrying particles are accelerated though a typical potential $\Delta \Phi = \gamma m_e c^2/e$ \citep{2009ApJ...703.1044B}, the dissipated power is 
 \be
\dot{U_B}  = I \Delta \Phi  = 2 \pi \gamma  \Delta \phi _{\rm max} {m_e c^3 \over e}r_{t,0} B_{NS} (r_{\rm max}/R_{NS})^{-3/2} = 3\times 10^{36} \, \gamma_3 b_{15} \left( { \Delta \phi _{\rm max} 
  \over 0.1} \right) \, \left( { r_{t,0} 
  \over 0.1 R_{NS} } \right) \, {\rm erg s}^{-1},
  \label{LX}
 \ee
 where we assumed that $r_{\rm max}= 2 R_{NS}$. This value is close to the observed persistent luminosity of magnetars. 

  For those flux tubes that satisfy $ t_{\rm res} \leq  t_H$, Eq. (\ref{ffff}), the twisting due to Hall drift  is relieved gradually via resistive untwisting. Estimating  $\Delta \phi \sim t_{\rm res} /t_H \leq 1$,  the dissipation rate is
  \be
  \dot{U_B}  = {c B_{NS}^3 R_{NS}^3 \over 4 e n L^2   } { r_{t,0}^2 \over r_{\rm max}^2} \approx {B_{NS}^2  R_{NS}^3 \over t_H}.
  \ee
 Note the strong dependence of the dissipated power on the \Bf, $\propto B^3$ (available energy $\propto B^2$, dissipation time scale  $\propto B^{-1}$). This explains why magnetar activity, especially giant flares, is mostly seen in the high field \NSs: in our model all \NSs\ show magnetar-like activity, but the continuously dissipated power has a cubic dependence on the \Bf. The scaling $L_X \propto  B^3$  is consistent with data \citep[Fig. 13 of][]{2014ApJS..212....6O}.
  Also major flares have a threshold \Bf, 
 Eq. (\ref{betaA}), while smaller flares can occur in lower field/older \NSs.

 \subsection{Flares: Hall twisting versus  resistive untwisting of  flux tubes}
 
 External twist is driven by the Hall evolution of the electron fluid within the crust and is  relieved via resistive dissipation of the external currents (which is linear in the  twist angle, Eq. (\ref{LX}). Thus,  the twist angle evolves according to
 \ba &&
  \Delta \dot{ \phi} = {1\over t_H} - { \Delta  \phi \over t_{\rm res}}
 \nn &&
 \Delta  \phi = {t_{\rm res} \over  t_H}  \left( 1- e^{-t/ t_{\rm res}} \right)
 \ea
 where the resistive time scale is 
 \be
 t_{\rm res} = { \Delta \phi  \over \partial _t \ln U_B} = { e B_{NS}  \over 2 \gamma m_e c^3} {  r_{t,0} R_{NS}^{3/2} \over \sqrt{ r_{\rm max}}}\approx
 300\,  b_{16} \gamma_3^{-1} \, {\rm yrs} .
 \ee
 
 In order to reach instability, $  \Delta \phi  \geq 1$, it is required that the resistive time be longer than the Hall time scale,
 \be
 {  t_{\rm res} \over t_H} = {B_{NS}^2  r_{NS}^{3/2} \over 8 \pi \gamma  m_e c^2 n r_{t,0}  \sqrt{r_{\rm max}}},
 \label{ffff}
 \ee
 where we assumed $L \sim r_{t,0}$.
 
 Note strong dependence of this ratio on the \Bf, $\propto B_{NS}^2$, on the size of evolving region $\propto r_{t,0}^{-1}$ and on the electron density $ \propto  n^{-1}$ (crustal depth where the driving occurs). 
 Since the size of the region is related to the energy dissipated in the flare, we expect that smaller \Bfs\ can induce smaller size flares, at shallower depths. This explains why magnetar activity is observed in  low field pulsars, but only  producing relatively  small burst and flares.  Major flares (like giant flares) that involve large fraction of the crust do require high \Bfs. 
 
 The condition for a magnetar to produce flares,  $ {  t_{\rm res}/ t_H} \geq 1$, can be expressed as 
 \be
 \beta_A = {v_A \over c} \geq  \sqrt{m_e \over m_p}  \sqrt{\gamma} \sqrt{Z\over A} { r_{t,0}^{1/2}  r_{\rm max}^{1/4} \over   R_{NS}^{3/4}}.
 \label{betaA}
 \ee
 The last terms being of the order of unity, the most important terms are $ \sqrt{m_e/ m_p}  \sqrt{\gamma}$, which evaluates to an order of unity. For $10^{16}$ G \Bf, the corresponding density is
 \be
 \rho \approx 10^{10} b_{16}^2 {\rm g cm}^{-3}.
 \ee

 Summarizing, smaller scale bursts and flares can be produced in smaller fields \NSs\ and/or at smaller crustal depth. (So that the twisting of external field by Hall drift occurs faster than the resistive untwisting  of the external currents). Giant flares, which require large scale re-configuration of the \ms\ require large \Bfs\ and shallower depths, where crustal density is $\rho \sim  10^{10}  {\rm g cm}^{-3}$.

 \section{Internal dissipation of magnetospheric \Alfven pulse during magnetar  flares}
 
 Observations of the post-flare evolution of the surface thermal emission indicate that in addition to external dissipation of energy, a considerable amount is also dissipated inside the \NS\ \citep{2006astro.ph.11747E,2012ApJ...761...66S}.
 In the Section we discuss how {\it external trigger can in addition  lead to internal dissipation}. Note, that this is opposite to what is assumed in the star-crack model, where internal trigger leads to external dissipation.
 
 In the ``Solar flare'' model of magnetospheric energy release the flares are associated with instabilities developing in the twisted current-carrying \ms, \eg\ when a  magnetic flux tube becomes unstable to kinks and reconnection. As a result  \Alfven waves are launched from the reconnection site down to the stellar surface. As the \Alfven shock  hits the surface it launched a whistler waves in the crust. On the Hall time scale of the 
crust such an impulse is nearly instantaneous. Below we demonstrate that the plastic dissipation of the elastic strain produced by such a pulse within  the crust is broadly consistent with  observations of the post-burst magnetar activity.

  \subsection{Interaction of \Alfven wave with the EMHD surface}
  
  Consider an \Alfven wave propagating along \Bf\ with amplitude $\delta B_A = B_0 k_A \xi_A$ ($k_A$ is a wave vector, $\xi_A$ is a displacement)  and hitting a \NS\ surface with plasma frequency $\om_p$. 
  First, for typical plasma parameters most of the  \Alfven pulse  is reflected. This can be seen from the Fresnel's reflection coefficients for the reflection of normal \EM\ at refractive index jump,
  \be
  R= \left( {n_1/n_2 -1 \over n_1/n_2 +1} \right).
  \ee
  For highly magnetized magnetospheric plasma $n_1 =v_A^{(0)}/c \approx 1$ ($v_A^{(0)}$ refers to the \Alfven velocity in the \ms), while inside the \NS\ the whistler modes have $n_2=n_w = c k \om_B /\om_p^2 \ll1$. In this limit the reflection coefficient is 
  $R \approx 1- 2 n_w \rightarrow 1$.

  Reflection of \Alfven pulse by the surface of a \NS\ launches a   whistlers pulse  in the crust with the same frequency as the incoming \Alfven wave, so that
  \be
  k_w = { \sqrt{ k_A v_A^{(0)}} \om_p \over c \sqrt{\om_B}} =  \sqrt{ \om^{(0)} \over \om_B} {\om_p \over c},
  \ee
  where $ \om^{(0) }= k_A v_A^{(0)}$ is  a typical frequency of the \Alfven waves in the \ms. Since $ \om^{(0)}\ll \om_B$, the whistler waves launched in the curst have wavelength
  much larger that the skin depth, $\lambda_w \gg c/\om_p$, consistent with the EMHD assumption. Note also, that the transmission coefficient 
  \be
  T = {2 v_A^{(0)}/c \over n_w + v_A^{(0)}/c} \approx 2.
  \ee
Thus, as   the \Alfven wave launched at the reconnection region propagates toward the star, it is mostly reflected at the surface, launching  a whistler pulse  in the crust. 
In the next section we consider the propagation of a whistler pulse in the crust.

\subsection{Green's function for whistlers}
Consider a half-space of plasma  $x>0$ described by  constant density  electron MHD, while the region $x<0$ obeys MHD equations.
Let the unperturbed \Bf\ $B_0$ be in the $x$ direction. Introducing the electron displacement $\xi_e$ we can write the \Bf\ perturbations in  the EMHD  regime as
\be
{\bf \delta B} = (  - i {\bf e}_y + {\bf e}_z) B_0 \xi_e' =  (  - i {\bf e}_y + {\bf e}_z) B_0 \epsilon_e,
\label{xie}
\ee
where the prime signifies the derivative with respect to $x$ and $\epsilon_e=  \xi_e' $ is the strain in the electron fluid.
The EMHD Eq. (\ref{main}) then gives the   non-relativistic  Schrodinger-type equation for the electron displacement
\be
 \partial_t \xi_e - i  {\om_B  c^2 \over \om_p^2}  \xi_e^{\prime \prime}
 =0.
 \label{Green}
 \ee
The solution to Eq. (\ref{Green}) is given by the Green's function of a free quantum-mechanical  particle, or, equivalently,  by the Green's function of the diffusion equation  with a complex diffusion coefficient:
 \be
 G_\xi = - \sqrt{2\over \pi}{ \om_p \cos  \left( {x^2  \om_p^2 \over 4 \om_B  c^2  t} \right) \over  c \sqrt{ \om_B t}}.
 \label{kk}
 \ee
 (We assumed that the displacement is non-vanishing on the boundary; otherwise there is also Green' function $\propto \sin {x^2  \om_p^2 /( 4 \om_B  c^2  t)}$.)
 This is the Green's function for whistler modes; it is normalized to $\int _0 ^\infty G_e  dx =1$; the minus sign chosen so that the strain, Eq. (\ref{strain}), is positive. 
  The key feature of Eq. (\ref{kk}) is the wave vector dependence of whistler modes: initial $\delta$-function has a broad spectral content, that gets spatially separated at later times/larger distances. Thus the whistler pulse becomes more coherent with distance.
 
 For a given initial displacement $\xi_0(x)$ the general solution is
 \be
 \xi_e = \int dx_i  G_e (x-x_i) \xi_0(x_i).
 \ee
 The corresponding  Green's function for the strain is 
 \be
G_e =G_\xi ' = {1\over \sqrt{2  \pi}} { x \sin {x^2  \om_p^2 \over 4 \om_B  c^2  t} \over  c^3 \om_B^{3/2} t^{3/2}}.
 \label{strain}
 \ee
 Note, that for a fixed time the amplitude of the strain fluctuations {\it increases with depth}.

  \subsection{Rotational discontinuity in EMHD}
 \label{const}
\begin{figure}[h!]
\includegraphics[width=.99\linewidth]{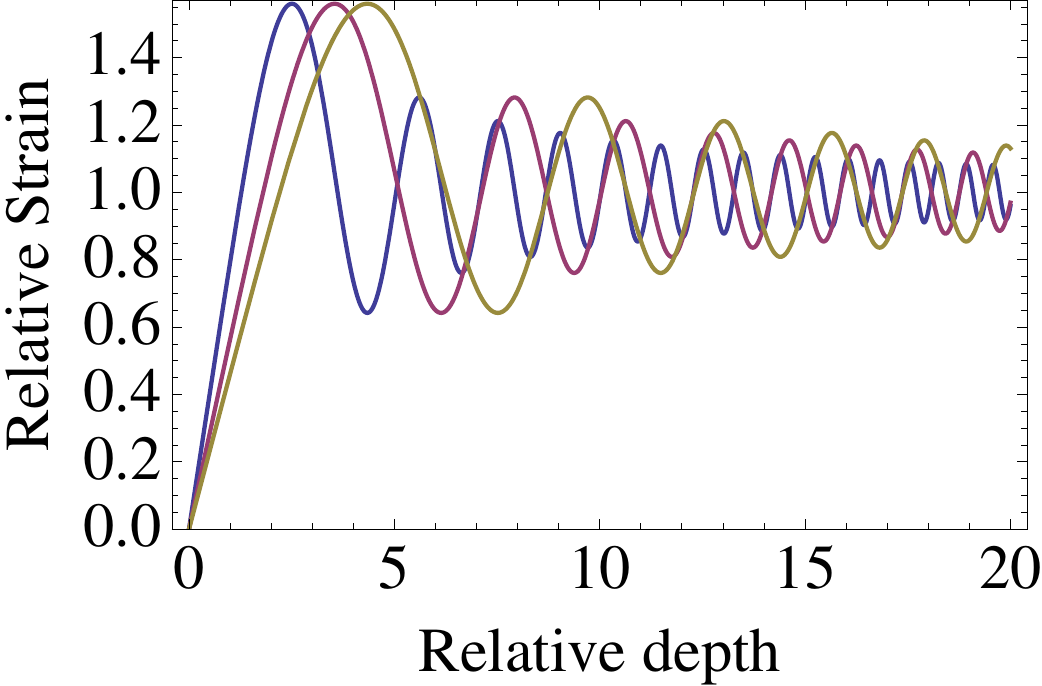}
\caption{Strain induced by Alfven shock in the crust. The plot shows a strain at three different times, illustrating evolution of the point of maximal strain to larger depths. }
\label{epsilonoft-shock}
\end{figure}

 A magnetar flare lasts  typically $\sim 100$ msec; this time is much shorter that the period; the resulting field disturbance can be treated as \Alfven rotational discontinuity. Next we consider  how such a structure interacts with the \NS\  surface. In an \Alfven shock \Bf\ in the plane of the discontinuity experiences a rotation by some angle. Thus, there is a surface  current density associated with it:
 \be
 {\bf g} =-  {c\over 4\pi} {[ \B ] \times{\bf e}_x},
 \ee
 where $[ \B ]$ is the jump in the transverse \Bf.
 The current pulse can be characterized by the transverse \Bf\ $B_\perp$ and  the rotation angle  $\alpha$, 
 \be
g  =  { c\over 2\pi}B_\perp    \sin (\alpha/2).
\ee
 In our representation of the \Bf\ in terms of the displacement vector, such initial condition corresponds to the discontinuous second derivative of $\xi_e$, or, equivalently, the  discontinuous first derivative of $\epsilon_e$.  The corresponding Green's function for the strain is 
 \be
 G_\epsilon ^{shock} = \int dx  G_\xi  = 2 {\cal C} \left( { x \om_p \over \sqrt{2 \pi} c \sqrt{ \om_B t}} \right),
 \ee
  where $ {\cal C} $ is the Fresnel integral of the second kind. (The superscript indicates that this is the Green's function for the Alfven shock). 
  In this case the strain in the electron fluid is  given by
 \be
 \epsilon_e = {  \sin (\alpha/2) } {\cal C} \left( { x \om_p \over \sqrt{2 \pi} c \sqrt{ \om_B t}} \right)  {B_\perp \over B_0},
 \label{epsilon}
\ee
see Fig. \ref{epsilonoft-shock}.

Note, that on the surface  $ {\cal C} (x=0)=0$. The maximum value of the strain is reached at $x = \sqrt{2 \pi} c \sqrt{ \om_B t}/\om_p$ and equals $2 {\cal C}(1) = 1.57$. Note that this value is independent of time.
Thus a whistler pulse propagates diffusively into the crust, conserving its maximal strain (in  constant density and assuming no dissipation - see discussion below where these constraints are relaxed).

The amplitude of the electron strain depends only on the properties of the initial \Alfven pulse, while the propagation depends on the properties on the medium. Thus,  in a varying density, in the WKB approximation, the strain is given by (\ref{epsilonee1}) with  $\om_ p$  and amplitude $\delta B$ depending on the depth, see \S \ref{dens}.

In time $t$ the whistler pulse reaches typical depth
\be 
x = 2 { c \sqrt{\om_B}  \over \om_p } \sqrt{t} \approx 100 \, {\rm meters} \, \rho_{10} ^{-1/2} b_{15}^{1/2}\, t_{day}^{1/2}.
\ee
This is approximately, the depth and the density at the outer-inner crust boundary,  close to the  neutron drip  point. Importantly, this is the depth where \cite{2012ApJ...761...66S} infer deposition of the energy during the outburst in Swift J1822.3-1606.

\subsection{Propagation of whistlers along  density gradient}
\label{dens}
In \S \ref{const} we considered propagation of whistlers in constant density, let us next take variations of density into account.
Consider \Bf\ along $z$ in  plasma with density changing along the field, $n = n_0 g(z)$. Let a whistler with frequency $\om$  propagate along $z$ direction.  Then 
\ba &&
\delta B_z =0
\nn 
&& 
\delta B_x = - i \partial_z \left( {\delta B_y ' \over g}\right) { \om_B c^2 \over \om_p^2 \om}
\ea
where $\om_p$ is defined in terms of $n_0$ and prime denotes derivative  with respect to $z$ coordinate.
The equation for $\delta B_y$ becomes
\be
\delta B_y =  { \om_B^2 c^4 \over \om_p^4 \om^2} \partial_z \left( {1\over g} \partial_z^2 \left( {\delta B_y ' \over g}  \right)\right),
\ee
which can be rewritten for ${\cal F}=   {\delta B_y ' / g}$,
\be
{\cal F} =  { \om_B^2 c^4 \over \om_p^4 \om^2} {1\over g}  \partial_z^2 \left( {{\cal F} ^{\prime \prime} \over g}  \right).
\label{gg}
\ee

We can solve this equation in a WKB-like approximation. Note, that the classical WKB is developed for the second order differential equations, while Eq. (\ref{gg}) is fourth order.
Rewriting (\ref{gg})  as
\be
{\cal F} =  { \om_B^2 c^4 \over \om_p^4 \om^2} {1\over g}  \partial_x^2 \left( {{\cal F} ^{\prime \prime} \over g}  \right) \epsilon^4, \epsilon \rightarrow 0,
\label{gg1}
\ee
we seek solution in the form ${\cal F} = e^{1/ \epsilon ( S_0 +  \epsilon S_1)}$. Expanding for $ \epsilon \rightarrow 0$, the zeroth order  gives
\ba && 
g^4 -  { \om_B^2 c^4 \over \om_p^4 \om^2}  {S_0^{\prime, 4}} =0
\nn &&
S_0 = \pm \sqrt{ \pm 1}  {\om_p \sqrt{\om} \over c \sqrt{ \om_B}} \int g(z) dz
\ea
and
\be
S_1 = - {1\over 4} \ln {g }.
\ee
Thus, 
\be
{\cal F}  \propto  g^{-1/4} e^{ i \int k_x (x) dx},
\label{F}
\ee
like in a classical WKB approach. Note that there are no resonances/reflection points.
Qualitatively,
\be
\om = {k(z)^2 c^2 \om_B \over \om_p(z)^2}.
\ee
Solution (\ref{F}) implies that $\delta B_y \propto g^{1/4}$. Thus, {\it  the amplitude of the whistlers  increases with depth}. This relation can be understood from the conservation of the Poynting flux $F_P$
carried by the wave:
\be 
F_p ={\om_B c^2 k \over \om_p^2}  \delta B^2 \propto {\sqrt{ \om_B \om} \over \om_p}  \delta B^2,
\label{dep}
\ee
from which it follows that $ \delta B \propto n^{1/4}$.

\subsection{Propagation of whistlers across density gradient}
\label{across}

\begin{figure}[h!]
\includegraphics[width=.99\linewidth]{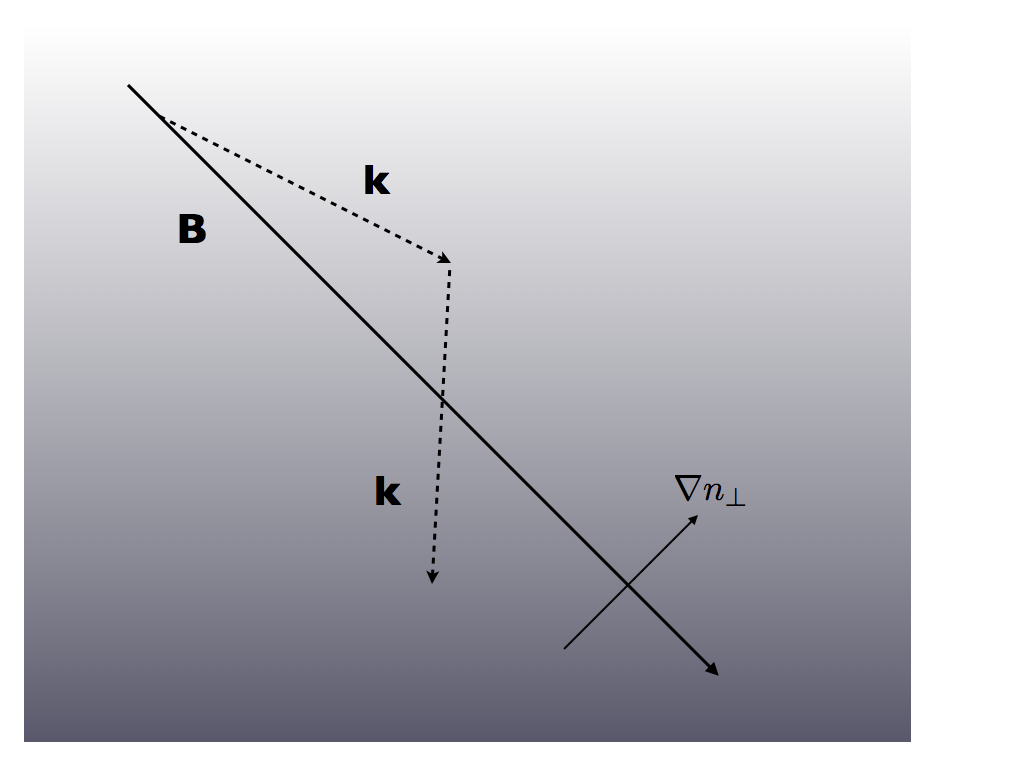}
\caption{Propagation of oblique whistlers in density gradient. If magnetic field is not aligned with the density gradient  whistlers are reflected from the high density regions in a direction of the component of the density gradient perpendicular to the \Bf. The shade indicates increasing electron density.}
\label{reflectionWhistler}
\end{figure}

Next consider \Bf\ along $z$ in a plasma with density changing in a transverse direction,  $n = n_0 g(x)$.  Let a whistler with frequency $\om$  propagate obliquely to the \Bf, so that $\delta B \propto e^{-i (\om t- k_x x)}$. We find
\ba && 
\delta B_y = - i {  g \om \om_p^2 \over \om_B c^2 k^2} \delta B_x
\nn && 
\delta B_z= i {\delta B_x'/k_z}
\nn &&
\delta B_x^{\prime \prime} = \left(k_z^2 - g^2 {\om_p^4 \om^2 \over \om_B^2 k_z^2 c^4} \right) \delta B_x.
\label{dd}
\ea
Equation for $\delta B_z$ has a form of the Schrodinger's equation with effective potential $V(z) \propto g^2$. Solutions of (\ref{dd}) are well known: they include propagating  and evanescent waves. For given $k_x$ and $\om$ the reflection occurs at $z_{\rm refl}$ given by 
\be
g(x_{\rm refl}) = {\om_B k_z^2 c^2 \over \om \om_p^2 }.
\ee
Thus, {\it oblique whistlers are reflected from the low density regions} in a direction of the component of the density gradient perpendicular to the \Bf, Fig. \ref{reflectionWhistler}.

\subsection{Whistler pulse with plastic response}

In the previous sections we considered  propagation of the whistler pulse within the crust neglecting  possible plastic response of the crust.  
If we take plasticity into account, the evolution will be approximately  described by (\ref{t}) with the driving term 
 \be
 \epsilon_e = {  \sin (\alpha/2) } {\cal C} \left( { z \om_{p,0}  \over \sqrt{2 \pi} c \sqrt{ \om_B t}} \sqrt{ n(z) \over n_0}  \right)  {\delta B \over B_0} \left({ n(z) \over n_0}\right)^{1/4}.
 \label{epsilonee1}
\ee
This expression for $\epsilon_e$ incorporates both the Green's fucntion in the constant density, Eq.  (\ref{epsilon}), as well as WKB amplitude changes with depth, Eq.  (\ref{dep}).

Similarly, the strain rate, 
  \be
\dot{ \epsilon_e } =  {  \sin (\alpha/2)\over \sqrt{2 \pi} }\, {z \om_p \over c t^{3/2} \sqrt{\om_B}}  \,\cos \left( { z \om_{p,0} \over \sqrt{2 \pi} c \sqrt{ \om_B t}} \right)  {\delta B \over B_0}  \left({ n(z) \over n_0}\right)^{3/4},
\label{kkk}
\ee
 sharply increases with depth due to: (i) increasing density and the correspondingly increasing  amplitude of the \Bf\ fluctuations (the  $  \left({ n(z) / n_0}\right)^{3/4})  $ factor  in Eq. (\ref{kkk}); (ii) 
dispersive effects, so that from an initial broad pulse the higher-frequency/higher wave number  components spatially separate (the  $z$ factor in Eq. (\ref{kkk})).

Overall, at time $t $ equal to the local Hall time $t_H$ the flow will reach a plasticity-mediated terminal value given approximately by  Eq.  (\ref{t}) with $t=t_H$,
 \be
 \epsilon_t \approx   {\epsilon_{crit} \over \ln \left({ c_s t_H   \over  d } { B_0 \over \delta B} \left( { n_0 \over n(z)} \right){1\over \sin (\alpha/2) \epsilon_{crit}^2} \right)}.
 \label{t00}
 \ee
The increasing density, the $ n_0/ n(z)$ term under logarithm, makes  the terminal $ \epsilon_t $ larger, contributing to higher rate of dissipation at larger depths, Eq. (\ref{W}). 

We conclude that an \Alfven pulse generated at the magnetospheric reconnection event launches a whistler pulse in the crust, that is effectively dissipated  around densities $\sim 10^{10}$ g cm$^{-3}$.

\section{Shear-density  EMHD instability  in \NS\ crusts}

In addition to linear evolution on Hall time scale (that builds crustal stress and leads to plastic deformation of the crust) and to the non-linear interaction of whistlers  \citep[that produces turbulent cascade to smaller scales, \eg][]{2013PhRvE..88e3103L},  the EMHD plasma is also  subject to a shear-density  EMHD instability  
\citep{2014arXiv1404.2145W} that requires (i) \Bf\ gradients on scales $L_B$; (ii) electron  density  gradient across the field $L_n$; (iii) density gradient should be on scales smaller than \Bf\ gradients, $L_n \leq L_B$. All these conditions are satisfied in the equatorial regions of \NS's crusts, where the \Bf\ is mostly in the $\theta$ direction, varying in the radial direction with electron density varying in radial direction as well. Scales of density variations are typically smaller than those of the \Bf:  below a total mass density of $10^{10} {\rm g \,cm}^{-3}$, the  density  drops to zero over about 100 meters.  

The unstable modes have a
length scale longer than the transverse density scale,
and a growth-rate of the order of the inverse Hall timescale.
Qualitatively, the instability is driven by the  current (velocity) shear that stretchers the perturbations and the rotation back by the whistler mode; the instability  also requires density gradient, making it somewhat analogous to the  magneto-buoyancy instability in regular MHD.

Next we consider shear-density  EMHD instability  in cylindrical geometry. This approximates the equatorial regions of the NS crust. 
 Consider cylindrical EMHD configuration with radially-dependent axial \Bf\  $\B_0 = f(r) {\bf e}_z B_0$, $n=n_0 g(r)$. For   axially-symmetric perturbations
 the EMHD equation (\ref{main}) gives
 \be
(r  \delta B_r ^{\prime} ) ^{\prime}  = 
\left(k_z^2 +{1\over r^2} + r \partial_r  \left( {f' \over r g} \right)  - {\om_p^4 \om^2 \over \om_B^2 k_z^2 c^4} \left( { g\over f}\right)^2 \right) \delta B_r .
\label{Eqcyl}
\ee

Multiplying by $r  \delta B_r ^\ast$ and partially integrating this gives
\be
\om^2 = {\om_B^2 c^2 k_z^2 \over \om_p^4}\,   {  \int (k_z^2 +1/r^2) | \delta B_r|^2  r dr   + \int | \delta B_r ^{\prime} - \delta B_r  \partial_r ( \ln f   ) |^2 r dr  - \int  \partial_r ( \ln f   ) \partial_r ( \ln g  )  | \delta B_r|^2  r dr  \over \int | \delta B_r|^2 \left( { f \over g}\right)^2  r dr }.
\label{inttt}
\ee
The instability can be driven by the last term in the numerator. The instability requires that $ f' g' >0$ - both \Bf\ and density decreasing with radius.  From Eq. (\ref{inttt}) it follows that the constant density configurations are stable, $\om^2 > 0$  since the last term is zero, while the first two are positively defined.
Also, comparing the density and \Bf\ gradients in the second and third term in   Eq. (\ref {inttt}), we conclude that for the third term to dominate it is required that density changes on scales smaller than \Bf, $L_n \leq L_B$ (the exact  conditions on $L_n/L_B$ that leads to instability are mathematically challenging to derive.)

The typical growth of the instability can be estimated from the last term in Eq. (\ref{inttt}) as the whistler travel time over the  scale $\sqrt{L_n L_B}$. 
 For further discussion of the shear-density instability in \NS\ crusts, see Appendix \ref{shearden}.

 \section{Discussion and Summary}

 In this paper we advance  a model of magnetar activity whereby  twisting of the external \Bfs\  by the electron Hall drift in the crust leads to both persistent emission as well as bursts and flares. In our model the elastic/plastic properties of the crust are not important for the production of flares: since the  \Bf\ in the crust is frozen in the electron fluid, and not the ion lattice, \Bf\ changes  can proceed without crustal faults, plastic or brittle. Evolution of the internal \Bfs\ drives  the electric currents into the \ms, which becomes susceptible  
to plasma instabilities.

We suggest that at low strain rates the response of the crust is mostly plastic and 
  present   a model of slow plastic deformations. 
  Employing a simple model of plastic deformations in metals, due to \cite{1960AuJPh..13..327G}, we found that the strain saturates at values well below the critical strain, Eq.  (\ref{t}). Importantly, the saturation level depends on the parameters of the problem only logarithmically, and thus remains  insensitive to the variations of these badly constrained parameters.

  The critical strain of metals is still an open equation. Ideal metals are expected to have a critical stress $\mu/(2\pi)$ \cite{Gilman}. Taking accounts of defects reduces the critical strain by approximately two orders of magnitudes,  \eg Peierls-Nabarro  critical stress \cite{1940PPS....52...34P,1947PPS....59..256N}.  In practice, critical stresses are another two orders of magnitude lower,  $\sim 10^{-4} \mu$. 
 Our result, that the plastically relaxing strain saturates at values $\sim 10^{-4}$ is, qualitatively,  consistent with this general scaling.

 The scaling (\ref{t}) also incorporates the fact that the behavior of a material depends on the rate of strain. For higher rates of strain (smaller $t_H$), 
 the saturation strain is higher, reaching, in the limit of very fast driving $t _H \sim b/c_s$, the critical strain. 
 Modern molecular dynamic numerical experiments \citep[\eg][]{2011arXiv1109.5095H}, that show a very high strain before development of a crustal faults, 
 occur in this vary fast regime due to computational limitations. As a results, we suggest,  they  cannot capture slow plastic deformation regime.

 While plastic recovery releases the crustal stress,  the \Bf, frozen in the electron fluid, undergoes twisting deformations, which  are pushed outside the \NS\ and are dissipated both continuously  and  in a Solar flare-like reconnection events.  These magnetospheric reconnection events launch \Alfven shocks that convert into fast high frequency whistlers upon hitting  the \NS\ surface. The resulting whistler pulse can propagate to the outer core  boundary and  dissipate the energy of the initial \Alfven pulse on a time scale of approximately a  day. This dissipated energy is detected as an enhanced post-flare surface emission.

We expect that {\it all \NSs\ show some kind of magnetar activity.} Plastic deformations dissipate magnetic energy, the rate of dissipation, Eq. (\ref{W}), is a cubic  function of the \Bf\ (two powers come from the energy density of the \Bf\ and one from the inverse scaling of the Hall time). For crustal \Bfs\ of the order of $10^{15} $ G the total dissipated power may reach $\sim 10^{34} $ erg s$^{-1}$. 

The persistent magnetar-like emission is unobservable in most rotationally powered \NSs. 
Giant flares, which  involve rearrangement of the \Bf\ on the scale of a whole \NS\ \ms,  can occur only  if the \Bf\ exceeds some threshold value, while smaller flares and persistent emission occurs in smaller  fields. Two competing effects are at play: larger scales EMHD deformation lead to increase twist of the external \Bf. On the other hand, non-linear interaction of 
whistlers lead to generation of smaller scales. Both processes occur, approximately, on Hall time scales $t_H$. 
 
 In addition to large scale whistler motions, the non-linear interaction of whistlers will lead to the   development of a turbulent cascade, generating smaller scales ion older pulsars. Adopting quasi-isotropic turbulent cascade, advocated by \cite{2013PhRvE..88e3103L}, 
the \Bf\  power at $l\sim 1/k$ scales  $ \propto k^{-2}$,  so that $B_k \propto B_0  (k_0/k) ^{1/2}$;  where $1/k_0$ is the outer scale of the turbulence; this gives  $\dot{E}_k  \propto B_0^3  k^{1/2} $ - in a turbulent cascade more energy is released on smaller scales. Also, the non-linear energy transfer accelerates toward small scales, 
$t_{k}  \propto B_k^{-2} k^{-2}  \propto k^{-1}$.  This explains weak magnetar-like activity in an older, low \Bf\ pulsar \citep{2012ApJ...754...27R}.
 
 We argued that \Bfs\ in the equatorial regions of \NS\ crusts are unstable to shear-density instability, which develops on whistler time scale of the upper crust. 
In a stark contrast to MHD, this   is an axially-symmetric  instability of purely poloidal \Bf\ --  there is no analogue in the MHD case.
The regions near the magnetic equator are preferred  both due to the development of the above mentioned shear-density instability, as well as higher \Bfs\ (and thus shorter time scales) in the equatorial regions \citep{2010MNRAS.402..345L,2013arXiv1305.2542R}. The fact that the magnetar activity occurs mostly near the magnetic equator, while the spin down rate measured the current flowing near the magnetic poles is related to the  highly complicated timing behavior during flares \citep{2014ApJ...784...37D}.

Magnetars share many properties (strength of \Bf\ first of all) with normal radio pulsars. Detection of radio emission from magnetars \citep{2005ApJ...632L..29H} and magnetar-like emission from the rotation-powered pulsar \citep{Fotis08} further blends this distinction. What  makes magnetars different?  We see two possibilities: first, 
  the initial configuration of the \Bf\ at the moment of crust freezing may be an important factor affecting magnetar activity. Numerically, stable  magnetic configuration of fluid stars were fond to vary from nearly dipolar to complicated non-axisymmtrical shapes \cite{2008MNRAS.386.1947B}.  During magnetar phase, the internal structure of \Bfs\ evolve toward the dipolar equilibrium; no toroidal field is required to remain inside a star, though we expect that some toroidal flux is trapped inside flux surfaces fully enclosed within a star. Since no analytical approximation to the structure of \Bf\ in fluid stars exists, we cannot estimate, \eg  the difference in magnetic energies between the initial and final states. 

Secondly, for a given poloidal fields  the crustal \Bfs\ can vary in a wide range of values  \citep{2013MNRAS.433.2445A}. Since magnetar activity is proportional to the thrid power of \Bf, mild variations of the \Bf\ within the crust can lead to large variations in the dissipated power.
 
  The paradigm of magnetar evolution described above assumes that \Bf\ is confined mostly to the crust of a \NS. This is consistent with the possibility that strong magnetic fields are created by neutrino-driven turbulence in proto-neutron stars, which is mostly effective in the outer layers \cite{1995PhR...256....5C,1994ApJ...435..339H}. 
 An alternative possibility is that magnetar activity is driven by the core of the \NS, which expels \Bf\ after becoming a type-I superconductor.

I would like to thank Andrei Beloborodov,  Andrew Cumming, Jeremy Heyl, Konstantinos Gourgouliatos, Victoria Kaspi, Yuri Levin, Mikhail Medvedev and Jay Melosh for discussions. 
  
\bibliographystyle{apj}
  \bibliography{/Users/maxim/Home/Research/BibTex}

\appendix

\section{Shear-density instability  in \NS\ crusts} 
 \label{shearden}
 
 To illustrate more clearly the shear-density instability  in \NS\ crusts here we first discuss a special case of the density and \Bf\ that analytically shows the instability and then comment on the special case when the poloidal \Bf\ becomes zero at some points.
 
 Eq. (\ref{Eqcyl}) has  a particularly simple solution, that demonstrates the instability, for the case of similar scaling of the \Bf\ and density
 \be
 f=g= r^{ 3/8 - r^2/(2 r_0^2) }.
 \ee
 In this case 
 \ba && 
 \delta B_r \propto \sin (k_r r)/\sqrt{r}
 \nn &&
 \om^2 = { \om_B^2 k_z^2 \over \om_p^4 r_0^2} \left((k_r^2 +k_z^2) -1/r_0^2 \right),
 \ea
 showing instability for $(k_r^2 +k_z^2) < 1/r_0^2$.

Two other special cases include 
 $f=g= e^{-r^2/( 2 r_0^2)}$, $\delta B_r = J_1 (k_r r)$, 
\be
\om^2 = { B_0^2 \over n_0^2} k_z^2 (k_r^2 + k_z^2),
\ee
and neutrally stable  case $f= (r/r_0)^{3/8} e^{-r^2/( 2 r_0^2)}$, $\om=0$. These solution give analytical  examples of the shearing density instability in the equatorial regions of the \NS\ crusts.

One mathematical subtlety concerns the convergence of integrals in (\ref{inttt}): it requires that the initial \Bf\ does not pass through zero. If $B_z =0$ at some radius,   $f(r=r_0) =0 $, then  expanding near the point $r=r_0$ gives
\be
\delta B_r ^{\prime \prime} = - {\om_p^4 \om^2 \over \om_B^2 k_z^2 c^4}  { g(r_0)^2  \over (r-r_0)^2}\delta B_r .
\ee
The requirement that perturbation vanishes at $r_0$ then leads to $\om=0$ - neutral stability. We expect that surfaces where $f=0$ will be unstable to resistive tearing instability \citep{2014arXiv1404.2145W}.

 \end{document}